\documentclass[reprint,aps,twocolumn,notitlepage,superscriptaddress]{revtex4-1}
\usepackage{graphicx}
\usepackage{amsmath}
\usepackage{amssymb}
\usepackage{csquotes}
\MakeOuterQuote{"}
\usepackage{bm}
\usepackage{overpic}
\usepackage{pict2e}
\usepackage{color}
\usepackage[capitalise]{cleveref}  
\usepackage[acronym]{glossaries-extra}
\usepackage{tikz}

\crefname{figure}{Fig.}{Figs.}
\crefname{equation}{Eq.}{Eqs.}
\crefname{section}{Sec.}{Secs.}

\glsdisablehyper
\setabbreviationstyle[acronym]{long-short}
\newacronym{re}{RE}{runaway electron}
\newacronym{mhd}{MHD}{magnetohydrodynamics}
\newacronym{pic}{PIC}{particle-in-cell}
\newacronym{iter}{ITER}{International Thermonuclear Experimental Reactor}
\newacronym{cfl}{CFL}{Courant–Friedrichs–Lewy}
\newacronym{rk4}{RK4}{4th order Runge-Kutta}
\newacronym{cpu}{CPU}{central processing unit}
\newacronym{gpu}{GPU}{graphics processing unit}
\newacronym{mpi}{MPI}{Message Passing Interface}
\newacronym{shm}{SHM}{MPI Shared Memory}
\newacronym{tae}{TAE}{toroidal Alfvén eigenmode}
\newacronym{rsae}{RSAE}{reversed shear Alfvén eigenmode}
\newacronym{gae}{GAE}{global Alfvén eigenmode}
\newacronym{cae}{CAE}{compressional Alfvén eigenmode}
\newacronym{flr}{FLR}{finite Larmor radius}
\newacronym{zlr}{ZLR}{zero Larmor radius}
\newacronym{fow}{FOW}{finite orbit width}
\newacronym{gs}{G-S}{Grad-Shafranov}
\newacronym{iaw}{IAW}{ion acoustic wave}
\newacronym{bae}{BAE}{beta-induced Alfvén eigenmode}
\newacronym{baae}{BAAE}{beta-induced Alfvén-acoustic eigenmode}
\newacronym{ae}{AE}{Alfvén eigenmode}
\newacronym{kaw}{KAW}{kinetic Alfvén wave}
\newacronym{lhd}{LHD}{Large Helical Device}
\newacronym{ttmp}{TTMP}{transit time magnetic pumping}

\begin{document}

\title{Self-consistent simulation of compressional Alfvén eigenmodes excited by runaway electrons}

\author{Chang Liu}
\email{cliu@pppl.gov}
\affiliation{Princeton Plasma Physics Laboratory, Princeton, NJ, USA}
\author{Andrey Lvovskiy}
\affiliation{General Atomics, San Diego, CA, USA}
\author{Carlos Paz-Soldan}
\affiliation{Columbia University, NY, USA}
\author{Stephen C. Jardin}
\affiliation{Princeton Plasma Physics Laboratory, Princeton, NJ, USA}
\author{Amitava~Bhattacharjee}
\affiliation{Princeton Plasma Physics Laboratory, Princeton, NJ, USA}
\affiliation{Princeton University, Princeton, NJ, USA}

\begin{abstract}

Alfvénic modes in the current quench (CQ) stage of the tokamak disruption have been observed in experiments. In DIII-D the excitation of these modes is associated with the presence of high-energy runaway electrons, and a strong mode excitation is often associated with the failure of RE plateau formation. In this work we present results of self-consistent kinetic-MHD simulations of RE-driven compressional Alfvén eigenmodes (CAEs) in DIII-D disruption scenarios, providing an explanation of the CQ modes.  Simulation results reveal that high energy trapped REs can have resonance with the Alfvén mode through their precession motion, and the resonance frequency is proportional to the energy of REs. The mode frequencies and their relationship with the RE energy are consistent with experimental observation. The perturbed magnetic fields from the modes can lead to spatial diffusion of runaway electrons including the nonresonant passing ones, thus providing the theoretical basis for a potential approach for runaway electron mitigation.

\end{abstract}

\maketitle

Tokamak disruption is one of the most important challenges to the success of magnetically confined thermonuclear fusion through tokamaks\cite{lehnen_disruptions_2015}. High-energy \gls{re} beams can be generated during a disruption event. These electrons can carry a large portion of pre-disruption magnetic energy through induction electric field acceleration, and cause severe damage to the plasma facing material in case of localized loss to the wall\cite{boozer_pivotal_2018}. A shattered pellet injection system has been introduced in ITER for disruption and \gls{re} mitigation, and several complimentary mitigation strategies are being actively investigated.

Although a high-energy \gls{re} beams is dangerous to the operation of tokamaks, it can also provide energy for destabilizing kinetic instabilities, which can lead to energy and spatial diffusion of the \glspl{re} and suppress their population growth. This idea is supported by disruption experiments in DIII-D\cite{lvovskiy_role_2018,lvovskiy_observation_2019} and ASDEX Upgrade\cite{heinrich_investigation_2021}, where magnetic field oscillations with frequencies in the MHz ranges have been detected during the current quench phase when an \gls{re} beam is present. New diagnostics have suggested that the mode has a dominant polarization of a compressional wave\cite{lvovskiy_parametric_2023}. In DIII-D, it is also observed that the strong excitation of current quench modes is often associated with failure of \gls{re} plateau formation, providing evidence for the dissipation of  \gls{re} beam by the Alfvén modes. Compared to the higher frequency whistler wave instabilities\cite{fulop_destabilization_2006} observed in the low-density flat-top Ohmic scenarios\cite{paz-soldan_spatiotemporal_2017,spong_first_2018,liu_role_2018},  Alfvén modes are less affected by electron-ion collisional damping which is strong in the disruption phase due to the low electron temperature\cite{liu_compressional_2021}.

In this paper we report the first self-consistent simulation of the excitation of Alfvén mode driven by high-energy runaway electrons. The kinetic-MHD code M3D-C1-K\cite{liu_hybrid_2022} was employed for the simulation, in which \glspl{re} were treated as kinetic particles and simulated using the \gls{pic} method, and their current was coupled into the \gls{mhd} equations. Similar simulation methods have been used in previous studies for the \glspl{cae} driven by energetic ions\cite{belova_nonlinear_2017}, and shear Alfvén waves driven by energetic electrons\cite{wang_simulation_2020}. The electrons are simulated following the gyrokinetic equations including the relativistic and finite-Larmor-radius (FLR) effects. Simulation results showed that the modes are driven by high-energy trapped runaway electrons\cite{liu_compressional_2021}, and the excited mode spectrum depends sensitively on the resonance between the mode frequency and the trapped \glspl{re} precession frequency. It was also verified in the nonlinear simulation that the perturbed magnetic fields of the excited modes can lead to spatial diffusion of \glspl{re} including the non-resonant ones.

The simulation was set up as follows. A plasma equilibrium constructed using EFIT from DIII-D shot 178631 at t=1250 ms was used, which is 20ms later than the initial Ar injection triggering the disruption\cite{lvovskiy_role_2018}. After the injection, the plasma is mostly composed of electrons and Ar$^{2+}$. The core electron density is $n_e=4\times 10^{20}$m$^{-3}$ and the electron temperature is around 5eV. A temperature-dependent Spitzer resistivity is used in the Ohm’s law to account for the resistive damping of the mode. Dirichlet boundary condition is used for the velocity and the perturbed magnetic fields. For runaway electrons, we use a tail distribution with a peak energy and collimated along the magnetic field with finite pitch angle width. This equilibrium distribution can be written as $f_0=G(p)H(\xi)$, where $G(p)$ is the distribution of the absolute value of momentum $p$ normalized to $m_e c$ ($m_e$ is the electron mass and $c$ is the speed of light),
\begin{equation}
	G(p)=\exp\left[-(p-p_0)^2/\Delta p\right]
\end{equation}
The energy distribution is inspired by the kinetic simulation result of hot-tail generation, which shows that the remnant of the Maxwellian tail can form a bump-on-tail distribution after thermal quench with a peak ($p_0$) and finite width ($\Delta p$). $H(\xi)$ is the distribution of cosine of \gls{re} pitch angle ($\xi=\cos\theta=p_\parallel/p$, $p_\parallel$ is the \gls{re} momentum along the local magnetic field),
\begin{equation}
	\label{eq:xi-distribution}
	H(\xi)=\frac{A}{2\sinh A}\exp\left[A\xi\right]
\end{equation}
Where $A(p)=2\hat{E}/(1+Z^*)\cdot p^2/\sqrt{p^2+1}$, $\hat{E}$ is the parallel electric field ($\sim 3$V/m from experiments) normalized to the relativistic \gls{re} critical electric field $E_c=n_e e^3\log\Lambda/4\pi\epsilon_0^2 m_0 c^2$. The width of the pitch angle distribution is determined by the balance between the electric field drag and the collisional pitch angle scattering\cite{aleynikov_stability_2015}. Here the value of $Z^*$ is defined by taking into account the partially-screening effect of argon, given that the fast electrons can penetrate screening of bounded electrons of partially ionized atoms and get deflected by the nuclei charge. The pitch angle width is then calculated as\cite{hesslow_effect_2017,hesslow_effect_2018}
\begin{align}
	Z^*=&Z_{eff}+\frac{1}{\ln\Lambda}\frac{n_{Ar}}{n_e}\left[\left(Z_{Ar}^2-Z_{eff}^2\right)\ln\left(\bar{a}_{Ar}p\right)\right.\nonumber\\
	&\left.-\frac{2}{3}\left(Z_{Ar}-Z_{eff}\right)^2\right]
\end{align}
where $Z_{Ar}=18$, $Z_{eff}=2$ and $\bar{a}_{Ar}$ is the effective ion size of Ar$^{2+}$.
$f_0$ is then used to calculate the particle weight evolution. Note that in this work we assume that $f_0$ does not change during the mode excitation. In other words, we ignore the further acceleration or pitch angle scattering of the \gls{re} beam, assuming that the timescale of these processes are longer compared to that of the mode excitation.

\begin{figure}
	\begin{center}
		\begin{overpic}[width=0.4\textwidth]{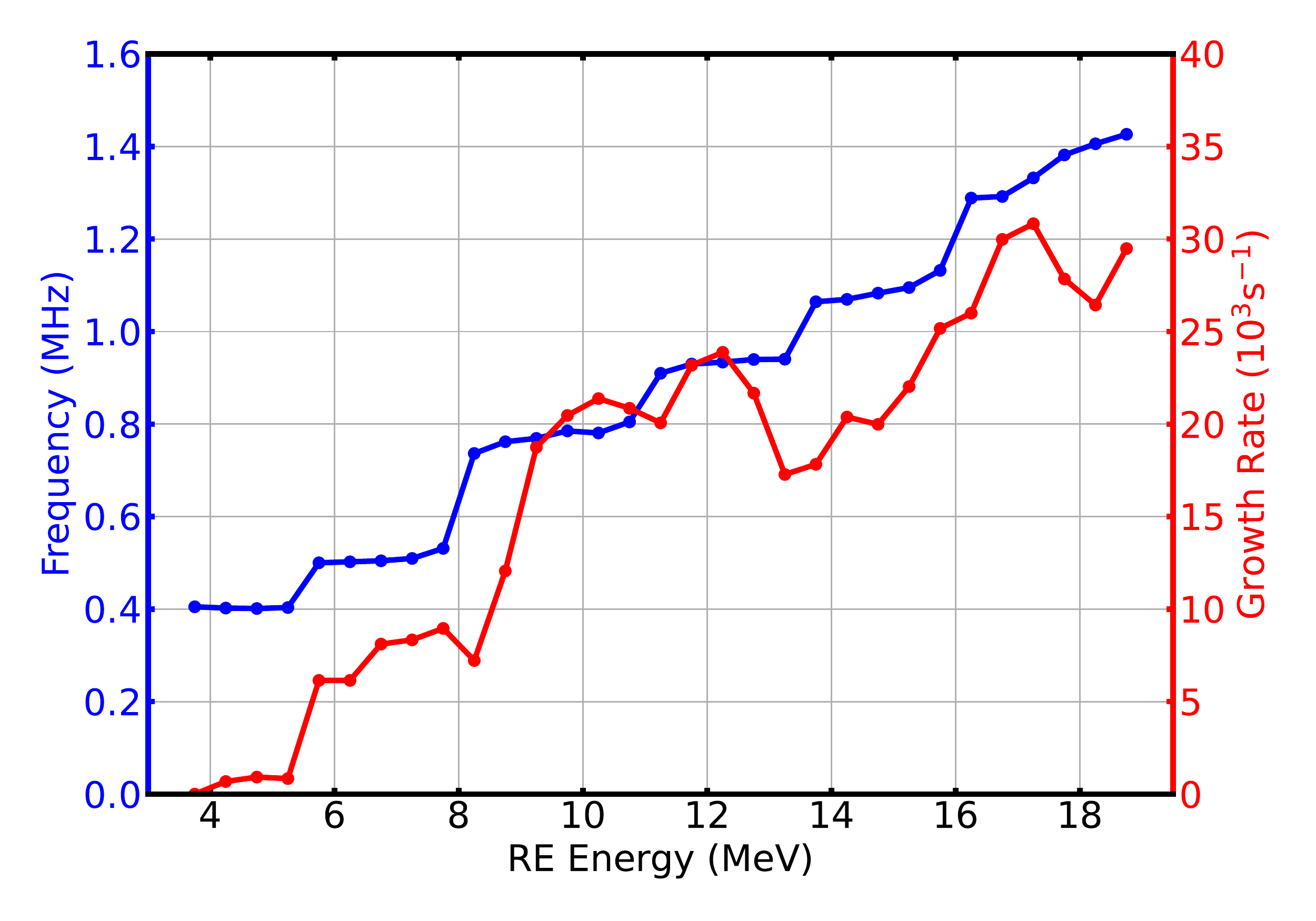}
			\put(29,28){\oval(13,5)}
			\put(55,42){\oval(15,5)}
			\put(80,12) {\textsf{(a)}}
		\end{overpic}
		\begin{overpic}[width=0.2\textwidth]{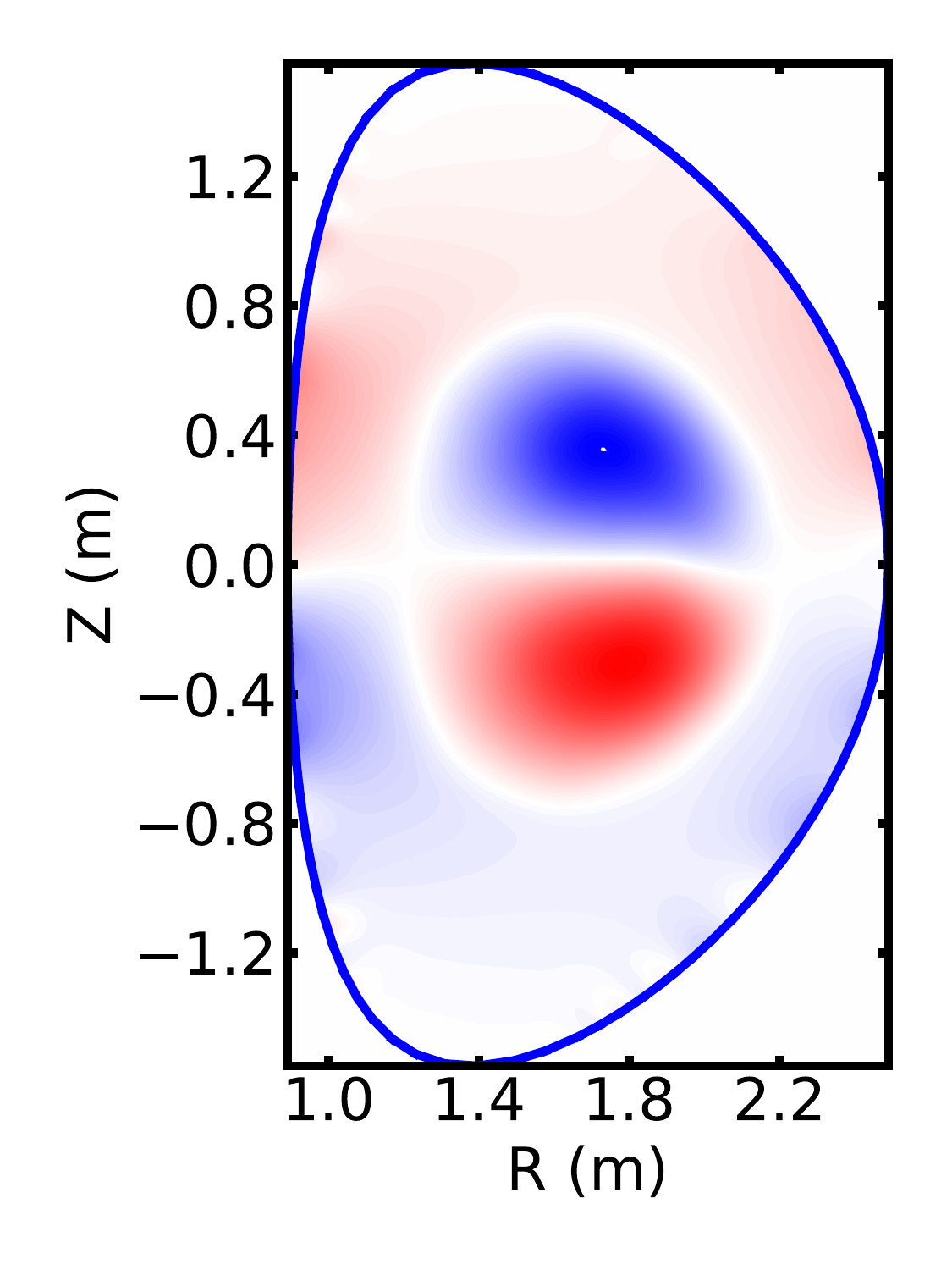}
				\put(60,19) {\textsf{(b)}}
		\end{overpic}
		\begin{overpic}[width=0.143\textwidth,trim=93 0 0 0,clip]{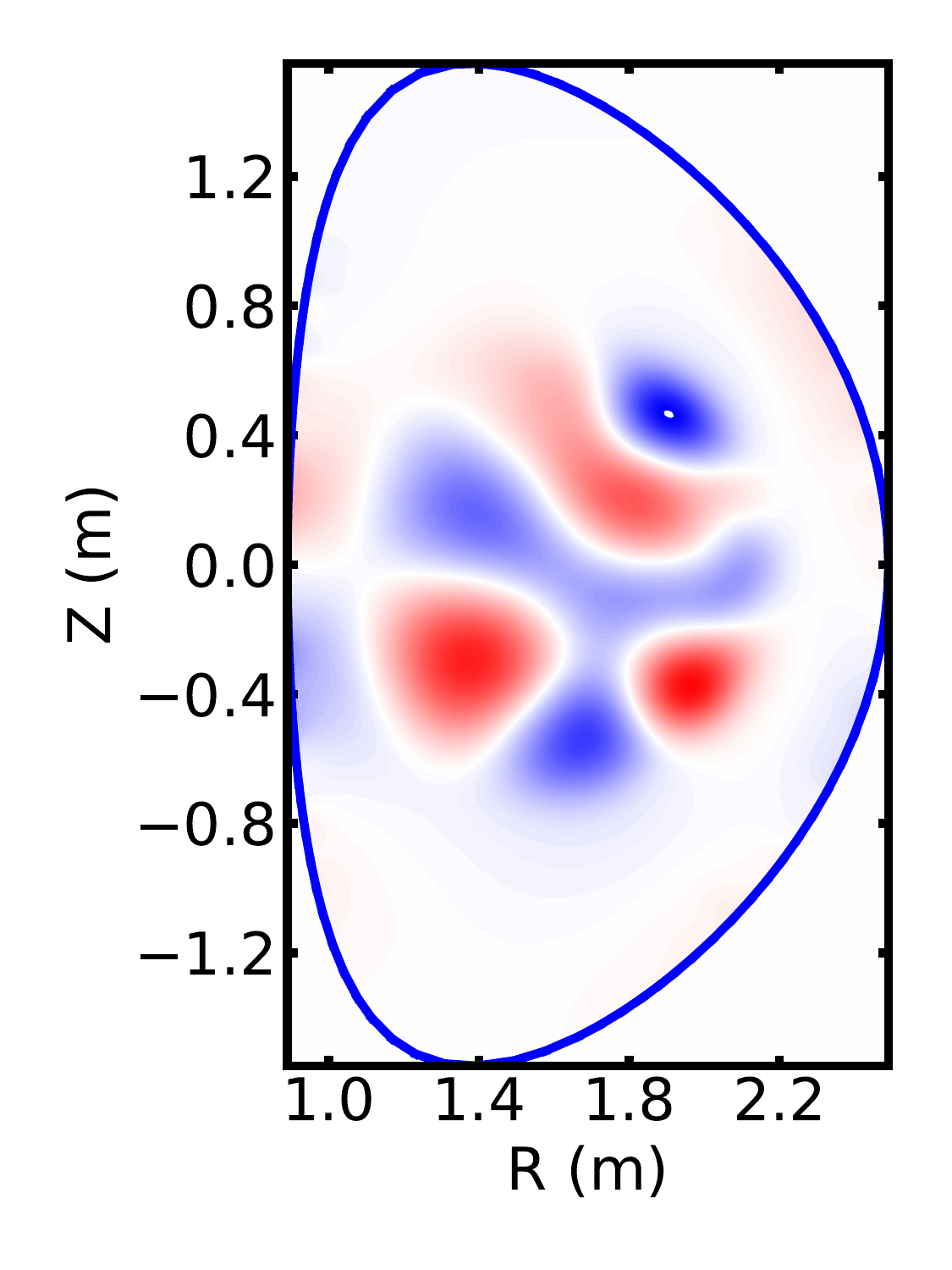}
			\put(40,19) {\textsf{(c)}}
		\end{overpic}
	\end{center}
	\caption{(a) Frequency (blue) and growth rate of dominant \gls{cae} from linear simulation using varying \gls{re} energy ($p_0$). The mode structure ($\delta B_\parallel$) of the two eigenmodes, corresponding to frequencies of 0.50MHz and 0.93MHz, are shown in (b) and (c).}
	\label{fig:frequencies}
\end{figure}

We first conducted linear simulations of \gls{cae} excitation. \cref{fig:frequencies} (a) summarizes the dominant AE frequency from linear simulations of $n=1$ mode using a narrow energy distribution $\delta p\sim 1.2$MeV and varying $p_0$ value. The core RE density is $3\times 10^{16}$m$^{-3}$. It is found that the frequency of the excited most unstable mode (blue line) follows a staircase-like function with \gls{re} energy, and the spacing between adjacent modes is about 0.2-0.4 MHz. The linear growth rates (red line) also have a positive correlation with \gls{re} energy with some fluctuations. The mode frequency magnitude and spacing between adjacent eigenmodes are consistent with experiments\cite{lvovskiy_role_2018}. However, the highest mode frequency we found is around 1.4 MHz, which is smaller than in the experiment. Each level of the staircase represents a different mode structure in the poloidal plane. The mode structures of 2 linear simulations are shown in \cref{fig:frequencies} (b) and (c). For all these excited modes, the perturbed magnetic fields satisfy $\delta B_\parallel\gg \delta B_\perp$, indicating that the mode is dominantly a compressional Alfvén eigenmode. Compared to shear Alfvén modes,  CAEs are global modes with large radial extent and can have components of multiple poloidal mode numbers. The eigenmode structure observed is consistent with the previous calculation of CAEs in NSTX\cite{fredrickson_non-linear_2013,smith_compressional_2017} by solving the Helmholtz equation.

\begin{figure}
	\begin{center}
		\begin{overpic}[width=0.22\textwidth]{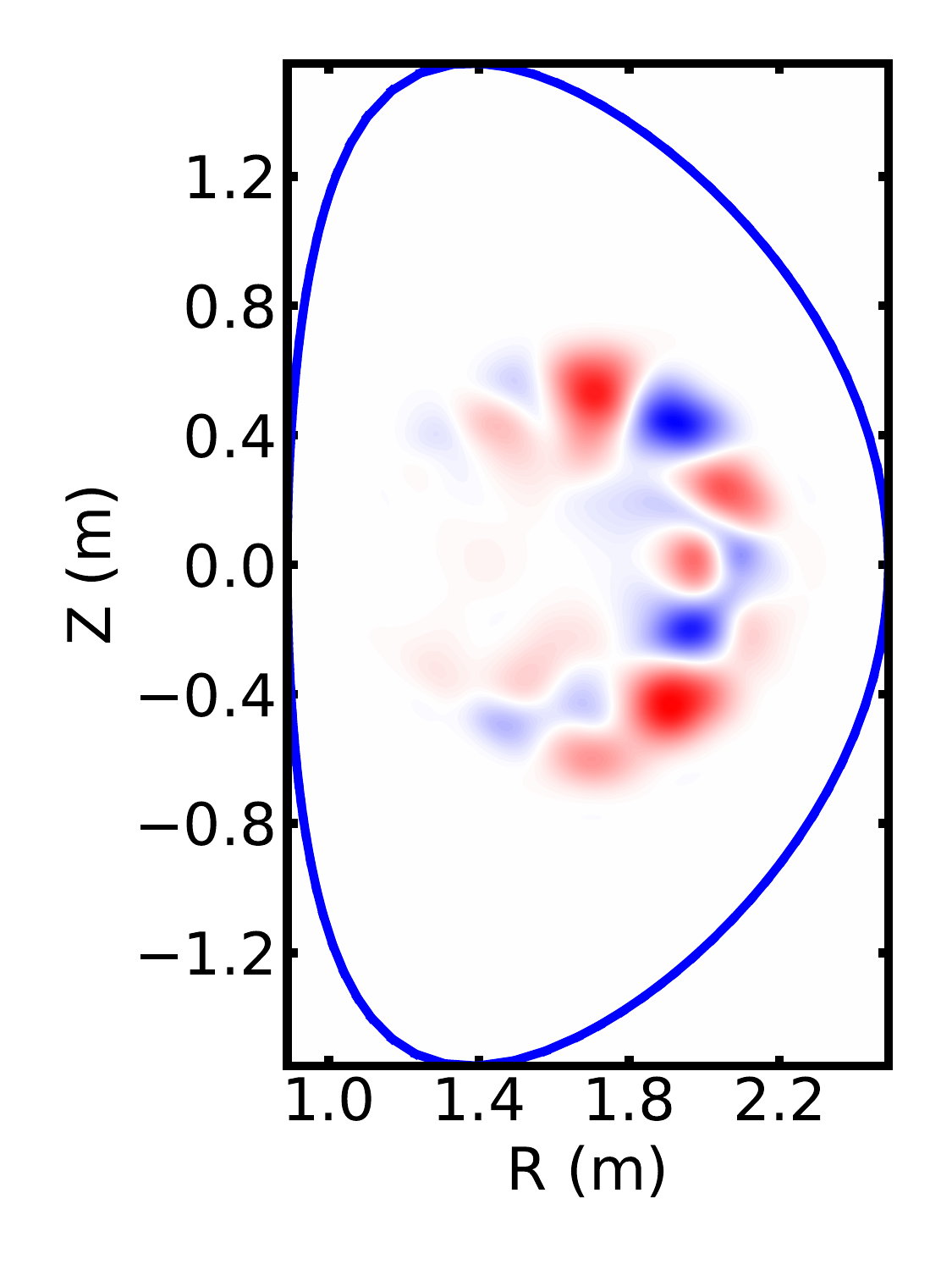}
			\put(62,20) {\textsf{(a)}}
		\end{overpic}
		\begin{overpic}[width=0.22\textwidth]{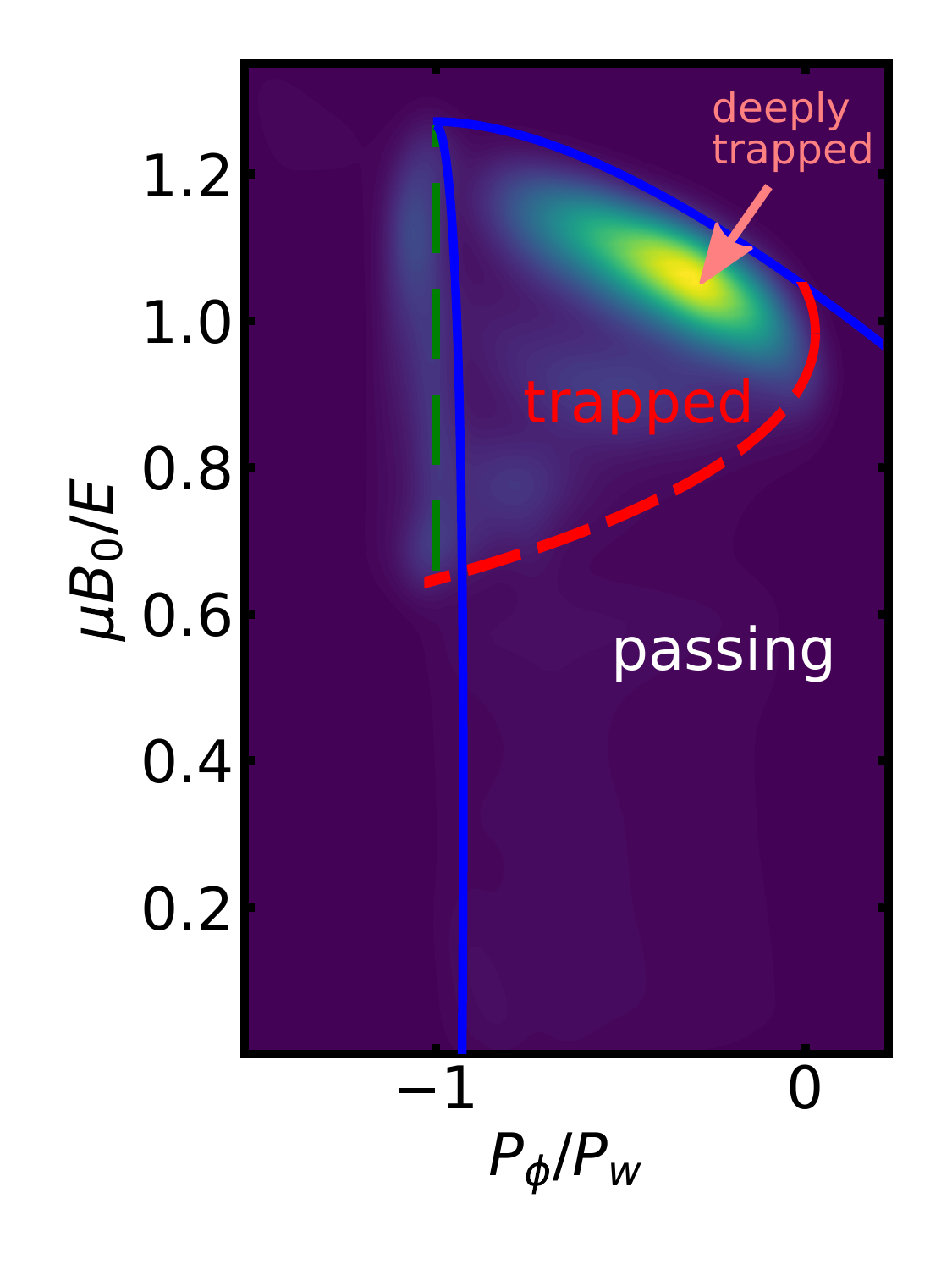}
			\put(62,20) {\color{white}\textsf{(b)}}
		\end{overpic}
	\end{center}
	\caption{(a) Distribution of \gls{re} $\delta f$ in the poloidal plane from linear simulation of 0.93MHz (\cref{fig:frequencies} (c)). (b) Distribution of absolution value of particle weight in the phase space of $P_\phi$ (toroidal momentum) and $\mu B_0/\mathcal{E}$ where $\mu$ is the RE magnetic moment and $\mathcal{E}$ is the particle kinetic energy. The boundary between trapped and passing particles is marked as red dashed line.}
	\label{fig:distribution-trapped}
\end{figure}

To understand the excitation mechanism of \glspl{cae}, the perturbation of the runaway electron distribution is analyzed. The distribution of particle weight for one linear simulation ($\omega=0.93$MHz) after the mode gets excited is shown in \cref{fig:distribution-trapped}, in which the boundary between trapped and passing electrons is marked. It is found that, although the \glspl{re} are mostly passing electrons, most of the resonant particles with large weight are deeply trapped electrons. This is because these trapped electrons can satisfy the resonance condition with the CAEs. For passing electrons, even considering the relativistic effect, both the cyclotron frequency ($\sim 5$GHz) and the transit frequency ($\sim 20$MHz) are too large compared to the mode frequency, and only a small population of barely passing electron can satisfy the resonance condition. However, for deeply trapped relativistic \glspl{re}, the toroidal precession frequency
\begin{equation}
	\omega_d=\frac{\gamma c^2}{r R \omega_{ce0}}
\end{equation}
can satisfy the resonance condition $\omega=n\omega_d$. Here $\omega_{ce0}=eB/m_e$ is the cyclotron frequency of unrelativistic electrons, and $R$ and $r$ are major and minor radius of trapped electron orbit. The precession motion is driven by the drift motion of \glspl{re} due to magnetic field curvature and gradient. Note that unlike the transit frequency which only depends on electron’s velocity, the precession frequency is proportional to the relativistic factor ($\gamma$). This resonance mechanism has been used to explain the excitation of \gls{bae} driven by supra-thermal electrons without considering the relativistic effect\cite{hl-2a_team_$ensuremathbeta$-induced_2010}. The resonance condition helps explain the linear relationship between the excited mode frequency and the runaway electron energy shown in \cref{fig:frequencies} (a).

For CAEs, most of the electromagnetic perturbation lies in $\delta B_\parallel$ and $\delta E_\perp$. The resonance electrons will be affected by the mirror force associated with perturbed field ($\nabla \delta B_\parallel$) and get pushed in the parallel direction, which is the \gls{ttmp} effect. The trapped electrons affected by the push have little gain or loss of parallel kinetic momentum due to the cancellation of energy exchange during the forward and backward motion. However, the force can lead to a change of canonical angular momentum, which can cause the trapped particle orbit to shift in the radial direction, similar to the Ware pinch effect. This means that a radial gradient of the trapped runaway electrons can provide a drive for the mode, similar to the energetic ion drive for the shear Alfvén mode. Following the derivations in \cite{liu_compressional_2021}, the growth rate of CAEs driven by \gls{re} can be estimated as
\begin{align}
	\label{eq:gammaL}
	\gamma_L=&\frac{4\pi^2 e^2}{\mathcal{E}_{mode}}\int\frac{|\langle G\rangle|^2}{\omega}\delta(\omega-n\omega_d)\times\left[\frac{\omega}{v_\parallel}\left(\frac{\partial}{\partial p_\parallel}\right)\right.\nonumber\\
	&\left.+\left(\frac{\omega R}{v_\parallel}-n\right)\left(\frac{\partial}{\partial \psi}\right)\right] f d^3\mathbf{p},
\end{align}
where $\mathcal{E}_{mode}$ is the total energy associated with the mode, $\langle G\rangle=\langle\delta\mathbf{E}\cdot\mathbf{v}\rangle$ describes the averaged energy exchange rate between the particles and the mode, $\psi$ is the poloidal magnetic flux. For trapped electrons with $v_\perp>v_\parallel$, $\langle G\rangle\approx \delta E_\perp v_\perp J_1(k_\perp \rho_e)$, where $\rho_e$ is the Larmor radius of the electron including the relativistic effect, and $J_1$ is the first-order Bessel function. Since $\rho_e\sim \gamma$ and $\mathcal{E}_{mode}\sim \langle \delta E_\perp^2\rangle$, the dominant contribution term ($\sim \partial/\partial \psi$) in \cref{eq:gammaL} is proportional to the radial gradient of resonant \glspl{re} and the resonant \gls{re} energy ($\gamma$). This means that given a fixed \gls{re} density and pitch angle distribution, higher energy \glspl{re} can provide stronger drive for the corresponding resonance mode. However, the pitch angle distribution of higher energy \glspl{re} is more peaked (\cref{eq:xi-distribution}), meaning that the population of trapped \glspl{re} is smaller. In addition, the collisional damping is stronger for higher frequency modes. These two factors limit the excitation of higher frequency CAEs. 

\begin{figure}
	\begin{center}
		\begin{overpic}[width=0.235\textwidth]{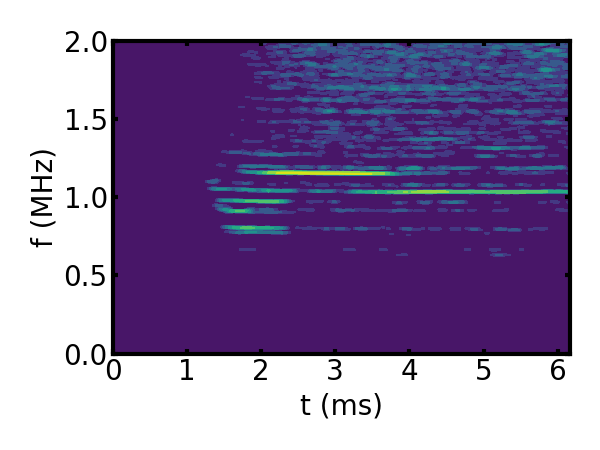}
			\put(37.5,16) {\tikz \draw[dashed,red] (0,0)--(0,2.2);}
			\put(85,18) {\color{white}\textsf{(a)}}
		\end{overpic}
		\begin{overpic}[width=0.235\textwidth]{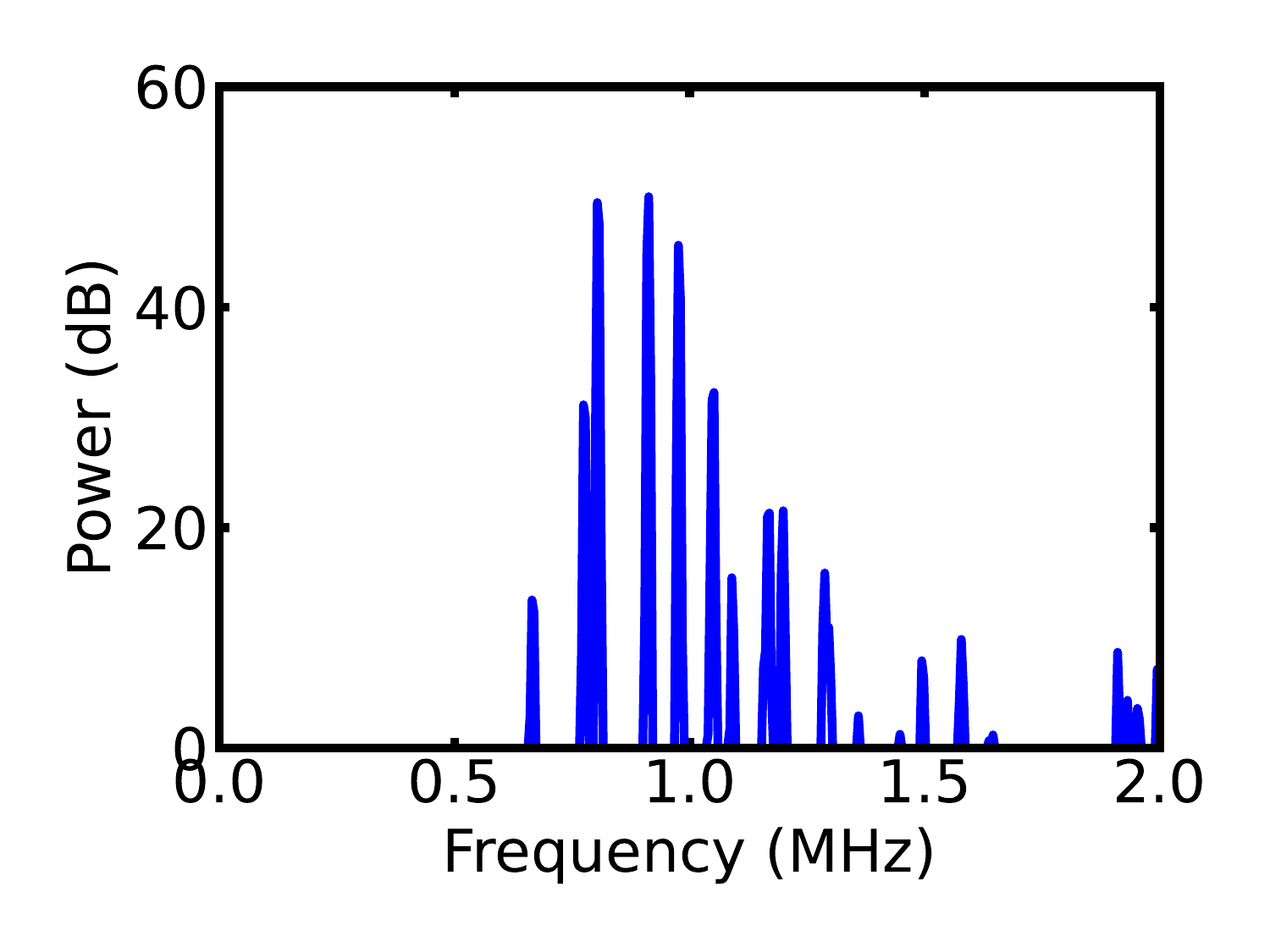}
			\put(80,18) {\textsf{(b)}}
		\end{overpic}
	\end{center}
	\caption{Spectrogram of excited CAEs from nonlinear simulation with only $n=1$ (a). The color indicated the mode power in the log scale. The power spectrum at $t=1.7$ms (red dashed line) is shown in (b).}
		\label{fig:spectro}
\end{figure}

Based on the linear simulation results, we conducted nonlinear simulation with $n=1$ mode only and simulations with multiple toroidal modes enabled. For $n=1$ simulation, the nonlinear terms in the PIC simulation are enabled, and the evolution of the \gls{re} distribution function will affect the mode saturation. A wider \gls{re} distribution in energy space ($p_0\sim 12$MeV and $\delta p\sim 7.5$MeV) is used. \cref{fig:spectro} shows the spectrogram of perturbed $\delta B_\parallel$ at the magnetic axis in one poloidal plane from the nonlinear simulation. It is found that multiple discrete modes can be excited and reach close amplitude at certain time (\cref{fig:spectro} (b)). In the later time of the simulation, some of the low-frequency excited modes decay in 1ms while other high-frequency modes can survive longer. The excited mode frequencies are almost constant in time without any frequency chirping. We also conducted full-torus nonlinear simulation including both the particle nonlinearity and coupling between different MHD modes. However, it is found that the excitation of the $n=0$ and $n=2$ components of magnetic perturbation is small in the simulation, indicating that the mode amplitude is not large enough to cause strong mode-mode interactions.

The simulation results are mostly consistent with the experimental observations \cite{lvovskiy_role_2018}. In experiments, it is found that the current quench mode can only be excited when \gls{re} energy passes certain threshold, and the low frequency modes are the first to get excited. This is consistent with our linear analysis as the excited mode frequency depends on the resonant \gls{re} energy. Note that in experiments, the hot-tail \glspl{re} are accelerated to higher energy by the inducted electric field, therefore the higher frequency modes will be driven later than the lower frequency ones. These multiple excited modes can coexist in the plasma like the nonlinear simulation results in \cref{fig:spectro}. Experimentally observed down-chirping of \gls{cae} frequencies is not observed in the nonlinear simulation, indicating the chirping may be caused by evolution of plasma parameters like ion density, rather than nonlinear effects. New diagnostic of the experiments\cite{lvovskiy_parametric_2023} reveals that the mode has a dominating compressional polarization, and the toroidal mode number is $n=1$, which are consistent with the nonlinear simulation. Note that $n=0$ modes are also identified in the experiments\cite{lvovskiy_parametric_2023}, but are absent in the simulations. This may be caused by the inaccuracy of diagnostics, or some other kinetic effects of \glspl{re} which can also leads to excitement of CAEs but are missing in our current simulation model.

\begin{figure}[t]
	\begin{center}
		\begin{overpic}[width=0.235\textwidth]{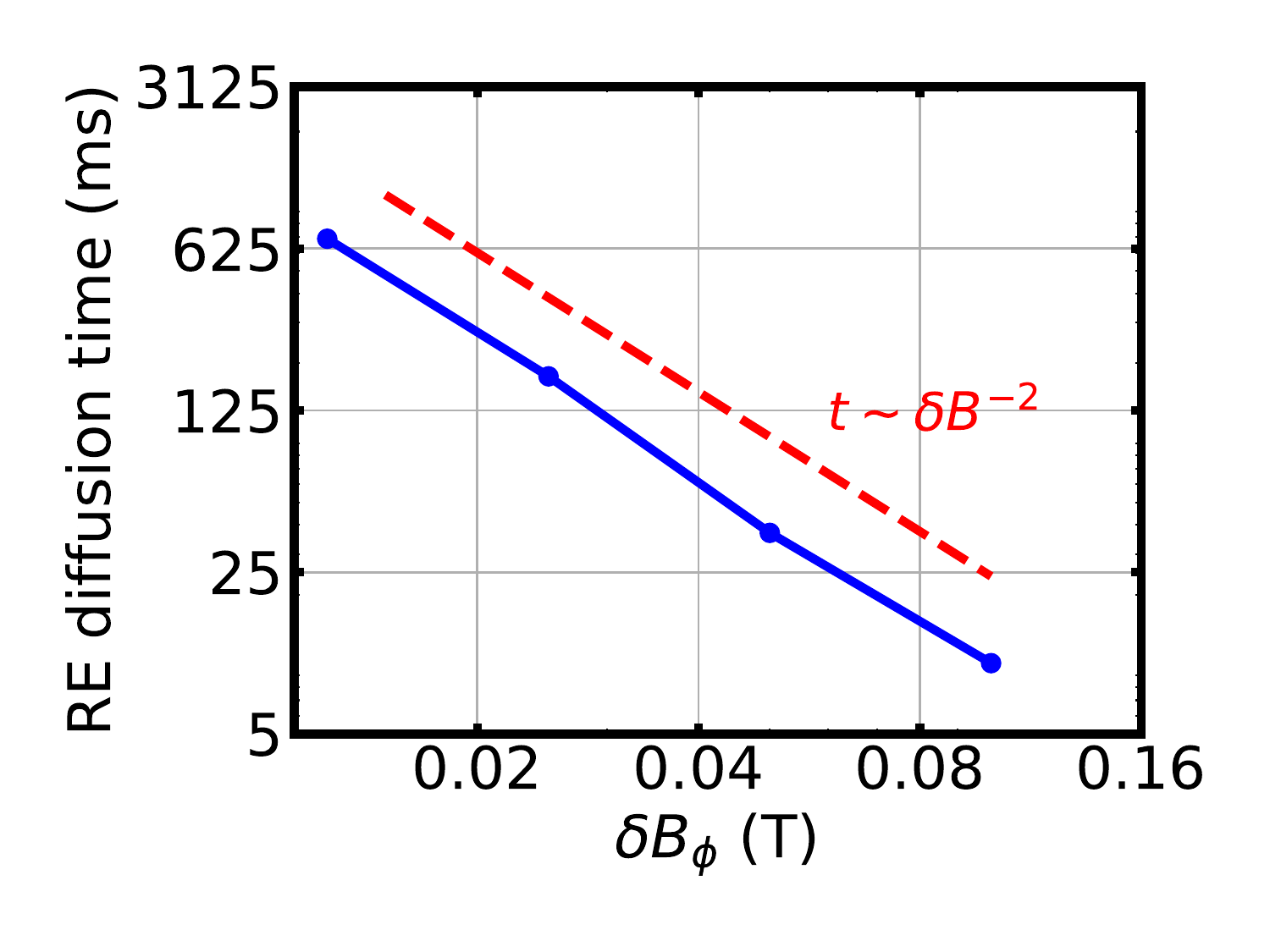}
			\put(79,20) {\textsf{(a)}}
		\end{overpic}
		\begin{overpic}[width=0.235\textwidth]{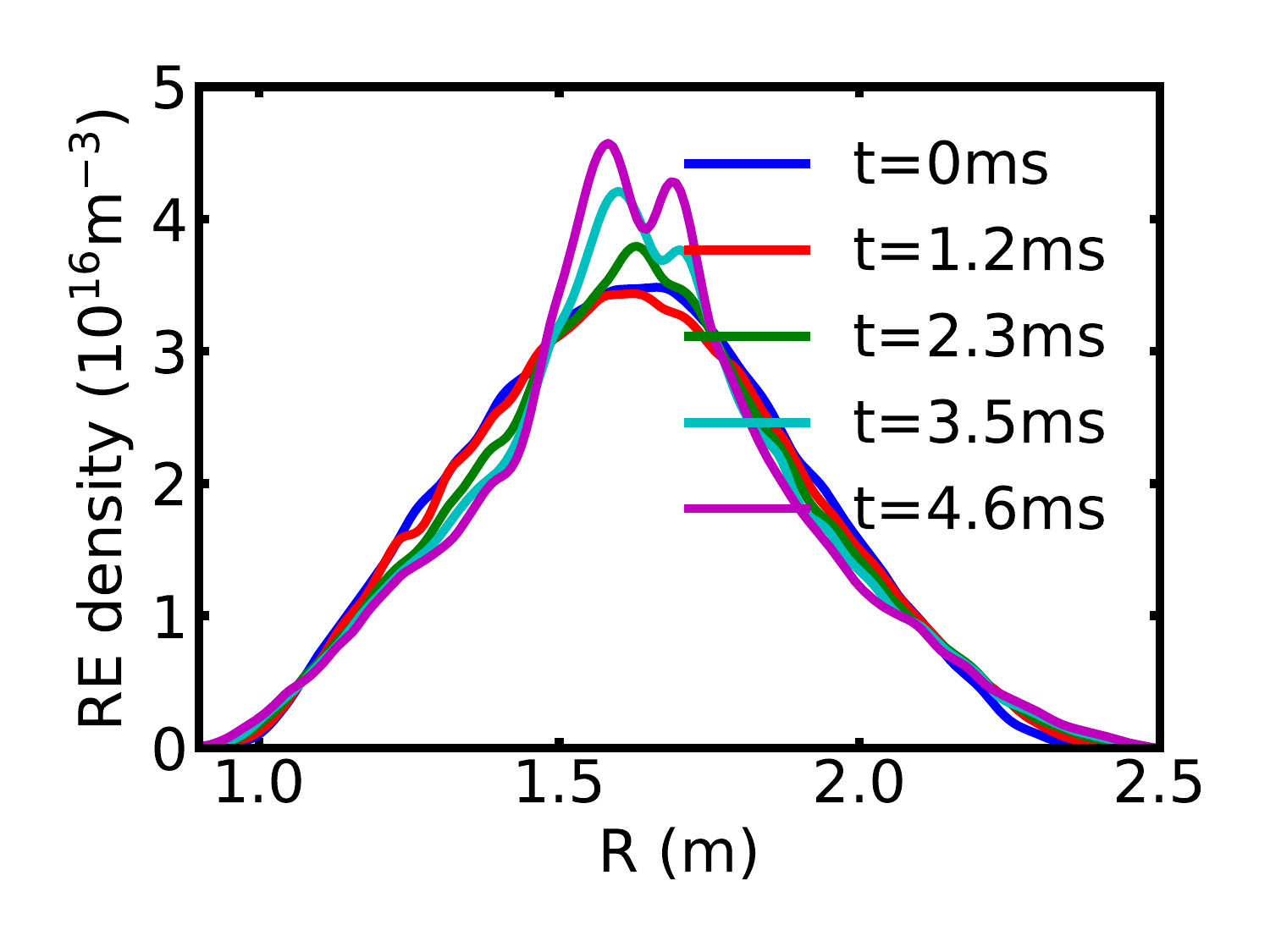}
			\put(79,20) {\textsf{(b)}}
		\end{overpic}
	\end{center}
	\caption{(a) \gls{re} diffusion time vs. $\delta B_\phi$ of CAEs, from nonlinear simulation with artificially amplified $\delta B$. The red dashed line showed the scaling law of $t\sim\delta B_\phi^{-2}$ (b) An example of \gls{re} density profile diffusion process captured after increasing the value of $\delta B_\phi$ to 0.05T.}
	\label{fig:diffusion}
\end{figure}

We further investigate the transport of \glspl{re} affected by the excited CAEs. For non-resonant passing \glspl{re}, a single mode can only lead to small deviation of particle orbit from its unperturbed one. But the overlapping of perturbed fields from multiple \glspl{cae} with different frequencies can cause decorrelation between particles and modes, and random walk of non-resonant passing \glspl{re} in plasma. Given that the random walk is mostly caused by the parallel motion with step size $c(\delta B_\perp/B_0) t_p$ ($t_p\approx R/c$ is the transit time of relativistic passing \gls{re} with zero pitch angle), and the decorrelation time is $1/\Delta f$ ($\Delta f$ is the spacing between adjacent \glspl{cae}), the diffusion time can be roughly estimated as
\begin{equation}
	\label{eq:tdiffusion}
	T_{\mathrm{diff}}\approx \left(\frac{a}{R}\right)^2\left(\frac{B_0}{\delta B_\perp}\right)^2\frac{1}{\Delta f}
\end{equation}
where $a/R$ is the inverse aspect ratio of the passing \gls{re} orbit. Following this estimation, to get a diffusion time around 10ms, the required value of $\delta B_\perp$ is about 0.012T, and the corresponding $\delta B_\parallel$ is around 0.12T. This value is larger compared to that in the nonlinear simulation. In order to test the passing \gls{re} diffusion, we rerun the nonlinear simulation starting from the point when multiple \glspl{cae} are excited (t=1.7ms), and artificially amplify the value of mode $\delta B$ to check the impact on \gls{re} diffusion.  The results of \gls{re} diffusion time are summarized in \cref{fig:diffusion} (a). The diffusion time is estimated by calculating the decaying rate of \gls{re} population inside the $r=0.3$m flux surface. An example of \gls{re} profile diffusion is shown in \cref{fig:diffusion} (b). It is found that although \glspl{re} in most of the region diffuse to the edge, the \gls{re} density near the magnetic axis increases. This may be caused by that diffusion near the axis is small, making it a stagnation point for passing \glspl{re}. The obtained diffusion time and its scaling law with $\delta B$ are consistent with the estimation in \cref{eq:tdiffusion}. However, according to the current simulation results, to reach such high amplitude of $\delta B$, the population of high-energy trapped \glspl{re} must be significantly larger than that in our current setup, which requires other pitch-angle scattering mechanisms than collisions.

In sum, in this paper we showed a self-consistent model to explain the current quench mode observed in DIII-D and other tokamaks, through kinetic MHD simulation with \glspl{re}. The excited modes are dominantly \glspl{cae}, which can have resonance with trapped \glspl{re} through \gls{ttmp}. The simulation results are mostly consistent with experimental observations, including the structure of the mode frequency spectrum, and mode polarization, and the relationship between the excited mode frequency and the \gls{re} energy. The amplitude of the excited modes from our simulation, however, is not large enough to drive significant loss of non-resonant passing \glspl{re}, which however has been observed in experiments.

The excitation of CAEs depends on the presence of trapped \glspl{re}. In our current model, these trapped \glspl{re} are generated from collisional pitch-angle scattering. Even though the partially screening effect of high-$Z$ impurities like argon is taken into account, the population of trapped \glspl{re} is still very small compared to the passing ones. It is possible that other pitch angle scattering mechanisms may also play important roles in the current quench phase, which can provide additional sources of trapped \glspl{re}. These mechanisms includes the scattering effect from higher frequency modes like whistler waves and high-frequency \glspl{cae}\cite{breizman_marginal_2023}, and scattering from turbulence fields\cite{martin-solis_effect_1999}. The diffusion of passing \glspl{re} driven by perturbed fields from \glspl{cae} is confirmed in the simulation, which shows the potential of this alternative \gls{re} mitigation strategy. Higher amplitude of $\delta B$ and stronger \gls{re} diffusion can be possibly achieved by launching the compressional wave using external coils\cite{guo_control_2018}, which which will be studied in future work.

\begin{acknowledgments}
Chang Liu would like to thank Elena Belova, Nikolai Gorelenkov and Neal Croker for fruitful discussion. This work was supported by the Simulation Center of electrons (SCREAM) SciDAC center by Office of Fusion Energy Science and Office of Advanced Scientific Computing of U. S. Department of Energy, under contract No. DE-SC0016268 and DE-AC02-09CH11466. This research used the high-performance computing cluster at at Princeton University, the computational resources of the National Energy Research Scientific Computing Center (NERSC) under Contract No. DE-AC02-05CH11231, and computational resources of the Oak Ridge Leadership Computing Facility (OLCF) under Contract No. DE-AC05-00OR22725. The data that support the findings of this study are available from the corresponding author upon reasonable request.
\end{acknowledgments}

\bibliography{paper}

\end{document}